\documentclass[a4paper]{jpconf}
\usepackage[backend=biber,style=ieee]{biblatex}
\usepackage{graphicx}
\usepackage{xcolor}
\usepackage{hyperref}
\usepackage{indentfirst}
\usepackage[ruled,vlined]{algorithm2e}

\include{pythonlisting}
\usepackage{lineno}

\begin{document}
\title{Variational Dropout Sparsification for Particle Identification speed-up}

\author{Artem Ryzhikov$^1$, Denis Derkach$^1$, Mikhail Hushchyn$^1$\newline on behalf of LHCb collaboration}

\address{$^1$ National Research University Higher School of Economics, 20 Myasnitskaya st., Moscow 101000, Russia}

\ead{aryzhikov@hse.ru}

\begin{abstract}
Accurate particle identification (PID) is one of the most important aspects of the LHCb experiment. Modern machine learning techniques such as neural networks (NNs) are efficiently applied to this problem and are integrated into the LHCb software. In this research, we discuss novel applications of neural network speed-up techniques to achieve faster PID in LHC upgrade conditions. We show that the best results are obtained using variational dropout sparsification, which provides a prediction (feedforward pass) speed increase of up to a factor of sixteen even when compared to a model with shallow networks.
\end{abstract}

\section{Introduction}

Particle identification (PID) algorithms play a crucial part in any high-energy physics analysis. A higher performance PID algorithm leads to a better background rejection and thus more precise results. Machine learning (ML) algorithms have gradually become the baseline approach for this task \cite{ml_pid}. One large family of such algorithms are neural networks. 

The main drawback of a deep neural network algorithm, however, is the time of prediction, which might become an issue in a high-load environment. This problem is particularly relevant in view of the forthcoming LHC upgrade, where the amount of collected data will be higher than ever. This work presents a study and comparison of modern speed-up techniques of neural networks applied to the PID problem. Techniques such as a full NN's configuration (like number of layers and neurons) search, pruning and variational dropout are considered and compared in the PID problem context.

\section{Problem statement}

The LHCb detector is a single-arm forward spectrometer covering the pseudorapidity range $2 < \eta < 5$, described in detail in Refs. \cite{LHCb}. Identification of various final state particles is performed by combining together the information from the LHCb detectors, namely from ring-imaging Cherenkov detectors (RICH), the electromagnetic and hadronic calorimeters, muon chambers (Figure \ref{fig:lhcb_detector}) and tracking system. Apart from the preaggregated likelihood such as observable subdetector responses \cite{likelihood_subd}, track geometry variables and different detector flags are also used. In addition to the presented solution, the muon identification \cite{muon_id} and calorimeter information about neutral clusters \cite{neutral_clusters} are also used. 

The PID algorithm objective is to identify the charged particle type associated with a given track. In the LHCb experiment there are five relevant particle species, namely, electron, muon, pion, kaon, proton, and ghost type (charged tracks that do not correspond to a real particle which passed through the detector) making a total of six hypotheses. Therefore, this is a multiclass classification problem. 

The aim of this research is to make PID algorithms \cite{ml_pid} faster. The research is focused on neural networks only.



\section{Existing methods}
In the following section we discuss several possible approaches to speed up the neural networks.

\subsection{Configuration grid search}
\label{ssec:gsearch}

One of the most commonly used methods to make neural network faster is finding its optimal configuration. Namely, an optimal number of layers and neurons of the neural network. Getting an optimal configuration of the neural network helps to find the necessary and sufficient complexity of the model for given data. It provides a good compromise between model speed and quality. However, such method has several drawbacks:
\begin{itemize}
    \item It requires a full search over all possible configurations. Even using advanced hyperparameter optimization techniques like \cite{GP} the search space is quite large.
    \item Due to the limited number of tested configurations the best configuration found is not the optimal one (in a global sense).
    \item The procedure is time consuming. Each tested configuration must be trained and evaluated.
    \item The procedure is not end-to-end. It requires multiple stages of training and evaluation instead of single one.
\end{itemize}

\subsection{Pruning}
\label{ssec:pruning}

Another commonly used and efficient family of techniques to improve feedforward performance of NNs is {\itshape neural network pruning}. Unlike the method from Section \ref{ssec:gsearch}, pruning is applied directly to a specific trained neural network instance. Namely, it is based on the idea of reducing the number of parameters during or after training. This approach makes it possible to train neural network only once, making the speeding up procedure much faster and more convenient. In this subsection we consider one of the most efficient pruning techniques to date \cite{L0Pruning, L1Pruning, TernaryQuantization}.

 The technique is called {\itshape quantization}. Originally it was based on the simple idea to move from high precision floating point types to lower precision ones. Moving to low precision reduces feedforward computation costs, making neural network faster. However, now there are lots of modifications of such a technique.

\begin{figure}
    \centering
    \includegraphics[scale=0.28]{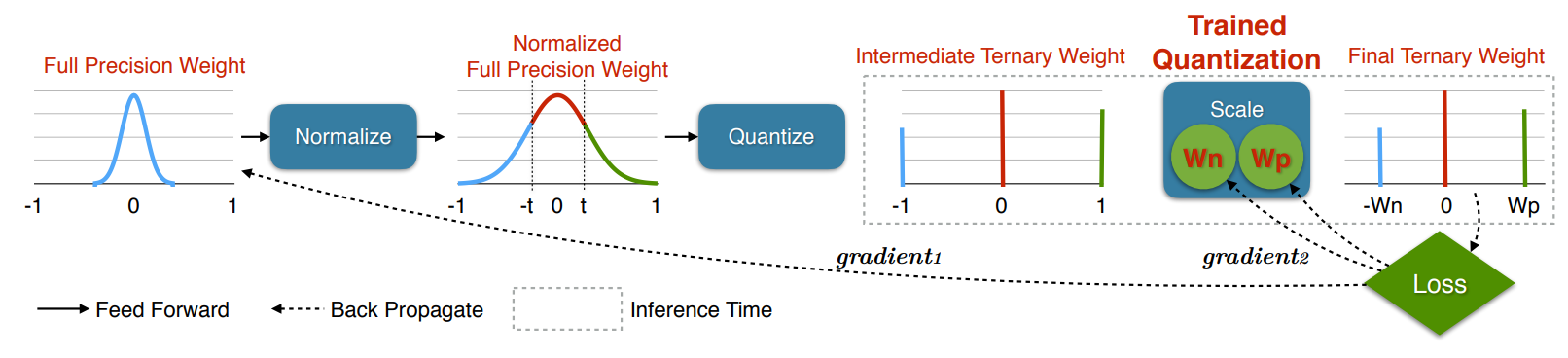}
    \caption{Trained ternary quantization \cite{TernaryQuantization}}
    \label{fig:trained_ternary_quantization}
\end{figure}

One such modification is {\itshape trained ternary quantization} \cite{TernaryQuantization}. It is based on the idea to move from individual parameter values to common ones. In \cite{TernaryQuantization} individual weights are replaced with one of three common values ($Wp$, $Wn$ and 0, Figure \ref{fig:trained_ternary_quantization}). Thus, the number of arithmetic operations in feedforward stage of the neural network can be reduced, making the neural network faster as well.

Since trained ternary quantization is a state-of-the-art \cite{TernaryQuantization} pruning technique, in this research we consider only this approach of pruning not taking into account another pruning techniques such as SVD and L-pruning \cite{L0Pruning, L1Pruning}.

\subsection{Variational Dropout}
\label{ssec:VarDrop}



\begin{figure}[!tbp]
  \centering
  \begin{minipage}[b]{0.4\textwidth}
  \centering
    \includegraphics[scale=0.25]{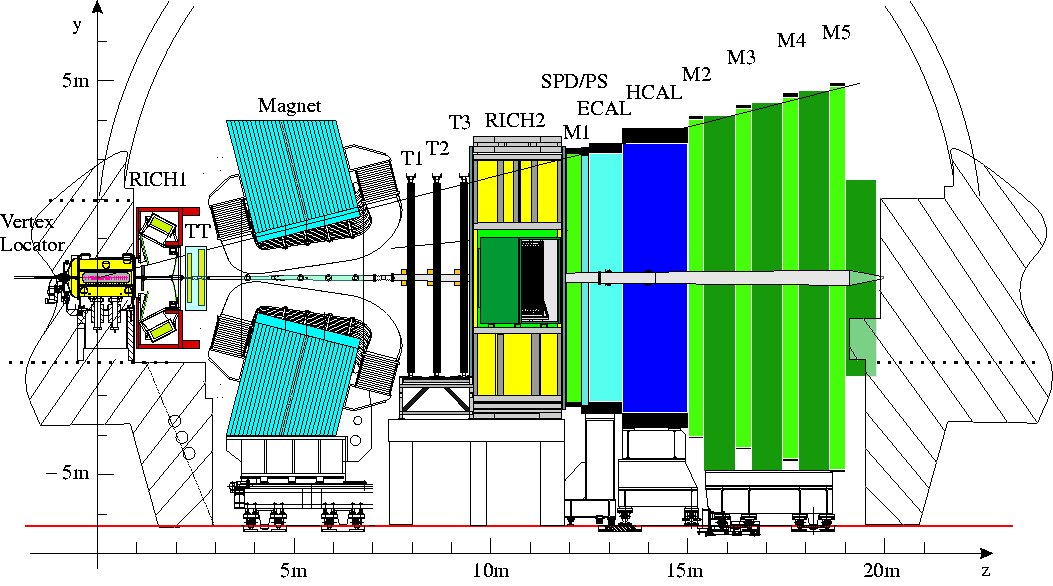}
    \caption{LHCb detector \cite{LHCb}}
    \label{fig:lhcb_detector}
  \end{minipage}
  \hfill
  \begin{minipage}[b]{0.5\textwidth}
    \centering
    \includegraphics[scale=0.27]{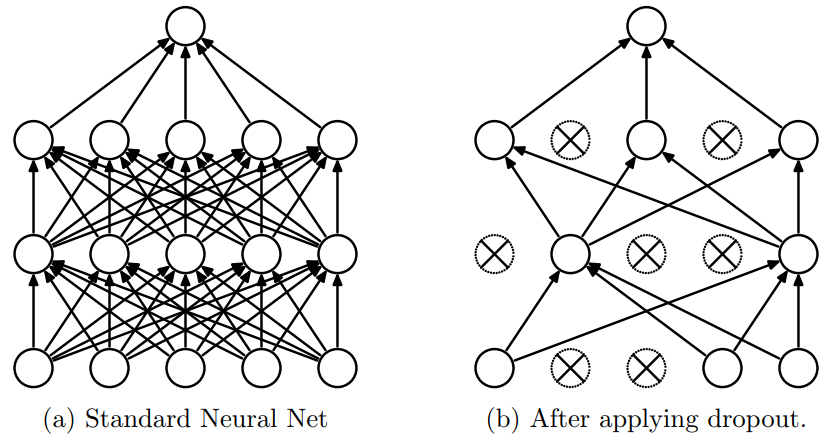}
    \caption{Dropout \cite{dropout}}
    \label{fig:dropout}
  \end{minipage}
\end{figure}

An alternative way to speed-up a neural network is to drop each parameter (zero connection's weight) separately with some probability $p$ (Figure \ref{fig:dropout}). Such a technique is quite common in deep learning and is called {\itshape dropout} \cite{dropout}. In practice dropout is a usefull technique which helps to prevent neural networks from overfitting. However, it requires the hyperparameter value $p$ to be defined. Moreover, each specific layer parameter is dropped (zeroed) randomly with the same probability $p$. It makes the original dropout implementation inappropriate for the automatic relevance determination (ARD) of neural network parameters, when all the redundant parameters are automatically dropped out during training stage. It makes it infeasible to sparsify a neural network effectively.

The authors of \cite{var_drop} propose an efficient and elegant way to train the dropout rate $p(\theta)$ for each trainable parameter $\theta$ in the whole range of possible values $\forall\theta: p(\theta)\in[0,1]$. The higher $p(\theta)$ for parameter $\theta$ the more likely for $\theta$ to be dropped (the less important $\theta$ is). Thus, such a technique helps to estimate the relevance for each parameter. The only thing remaining after training is to drop such a parameters $\theta$, whose dropout rate of $p(\theta)$ is close to 1. In this way, we can perform a speed up of the neural network.

\section{Data}
\label{sec:data}

In the simulation, pp collisions are generated using Pythia \cite{pythia} with a specific LHCb configuration \cite{lhcb_sim}. Decays of hadronic particles are described by EvtGen \cite{evtgen}, in which final-state radiation is generated using Photos \cite{photos}. The interaction of the generated particles with the detector, and its response, are implemented using the Geant4 toolkit \cite{geant4} as described in Ref. \cite{lhcb_sim_2}. 

The PID algorithms are trained on simulated samples with the 6 labeled particle types. The training sample is obtained from abundant simulated decays of heavy hadrons that emulate the kinematic distributions of signal samples studied in various LHCb analyses. Aggregated information from the LHCb sub-detectors, geometry, track reconstruction quality and kinematic properties are used as input features for the algorithms \cite{features}. Only long tracks are considered, which pass through both VELO, trackers and the calorimeter. The reconstruction quality of such tracks is highest and they are used in most LHCb analyses.

The experimental data consists of 6 million tracks (1 million trackes per each particle type). 50 \% of with were taken for train, 50 \% for test. Each sample (track) has 59 features.

\section{Results}

The quality of a model is measured by ROC AUC metric. Thus, the benchmark of the research is the model prediction speed at given ROC AUC (the ROC AUC of the baseline).

We implement all techniques described above to test them in the PID problem at LHCb. The results are presented in table \ref{tab:results}.

\begin{table}[ht!]
\scalebox{1.0}{
\lineup
\begin{tabular}{ll|llllll|l}
\br
&&\multicolumn{6}{c|}{\textbf{ROC AUC}}& \\

Method  & \# Neurons & Electron                               & Ghost  & Kaon                                   & Muon                                   & Pion                                   & Proton                                 & \textbf{Speed-Up}                            \\
\mr
6xDNN   & 45-48      & 0.9855                                 & 0.9485 & 0.9148                                 & 0.9844                                 & 0.9346                                 & 0.9178                                 & x1                                  \\
1xDNN   & 150        & 0.9863                                 & 0.9570 & 0.9145                                 & 0.9889                                 & 0.9463                                 & 0.9167                                 & x1                                  \\
Grid Search   & 30         & 0.9871                                 & 0.9557 & 0.9158                                 & 0.9893                                 & 0.9427                                 & 0.9125                                 & x5                                  \\
Pruning & Auto       & 0.9843                                 & 0.9435 & 0.9154                                 & 0.9834                                 & 0.9352                                 & 0.9110                                 & x5                                  \\
VarDropout    & Auto       & {\color[HTML]{009901} \textbf{0.9881}} & 0.9548 & {\color[HTML]{009901} \textbf{0.9244}} & {\color[HTML]{009901} \textbf{0.9896}} & {\color[HTML]{009901} \textbf{0.9509}} & {\color[HTML]{009901} \textbf{0.9228}} & {\color[HTML]{009901} \textbf{x16}} \\ 

\br
\end{tabular}
}
\caption{Performance of different methods}
\label{tab:results}
\end{table}



First two lines contain two equivalent baseline solutions for the PID problem \cite{features}. 

The first line corresponds to the baseline algorithm of 6 binary classifiers, where each classifier is a dense neural network with single hidden layer. 

The second line corresponds to the alternative baseline of single neural network with the same input features (Sec. \ref{sec:data}), single hidden layer and 6 outputs (number of classes). The size of hidden layer was chosen to be 150 neurons to make the number of parameters and inference time close to the original (first) baseline. 

The third line corresponds to the best configuration of the neural network provided by a full configuration search (grid search, Sec. \ref{ssec:gsearch}).  This approach provided a relative speed up of a factor 5 without loss of quality. However, it took lots of time to test all candidate configurations of the neural network to choose the optimal one.

The fourth line corresponds to one of the state-of-the-art pruning techniques - trained ternary quantization (Sec. \ref{ssec:pruning}).  It also provides a factor $5$ speed-up. However, the best configuration is found much faster. The neural network was trained only once with only the initial configuration. However, this approach lead to a significant loss of quality.

Finally, the last line corresponds to the ARD variational dropout solution (Sec. \ref{ssec:VarDrop}). It made the neural network approximately $16$ times faster without any loss of quality. Moreover, the neural network was trained in the end-to-end mode. Namely, it was trained only once with only the initial configuration of layers.

All the benchmarks were performed on CPU. All the neural networks were trained using {\it PyTorch} framework \cite{pytorch}.

\section{Conclusion}

Neural network speed up is a problem in a wide range of applications. In this research the most used speed up techniques were studied and compared in the application to the PID problem. The results shows that Variational Dropout Sparsification technique \cite{var_drop} provides the best results for the given problem. It speed up the PID neural network 16 times without any loss of quality.

The source code is available at\footnote{\url{https://github.com/HolyBayes/pytorch_ard}}

\section*{Acknowledgement}
The research leading to these results has received funding from Russian Science Foundation under grant agreement n° 17-72-20127.


\nocite{python}
\nocite{pytorch}

\printbibliography

\end{document}